\documentstyle[12pt]{article}

\font\twlgot =eufm10 scaled \magstep1
\font\egtgot =eufm8
\font\sevgot =eufm7
\font\twlmsb =msbm10 scaled \magstep1
\font\egtmsb =msbm8
\font\sevmsb =msbm7

\newfam\gotfam
\def\pgot{\fam\gotfam\twlgot}
\textfont\gotfam\twlgot
\scriptfont\gotfam\egtgot
\scriptscriptfont\gotfam\sevgot
\def\got{\protect\pgot}
\newfam\msbfam
\textfont\msbfam\twlmsb
\scriptfont\msbfam\egtmsb
\scriptscriptfont\msbfam\sevmsb
\def\Bbb{\protect\pBbb}
\def\pBbb{\relax\ifmmode\expandafter\Bb\else\typeout{You cann't use
Bbb in text mode}\fi}
\def\Bb #1{{\fam\msbfam\relax#1}}

\newcommand{\cG}{{\got g}}

\let\Large=\large

\def\op#1{\mathop{\fam0 #1}\limits}

\newcommand{\beq}{\begin{equation}}
\newcommand{\eeq}{\end{equation}}
\newcommand{\ben}{\begin{eqnarray}}
\newcommand{\een}{\end{eqnarray}}
\newcommand{\be}{\begin{eqnarray*}}
\newcommand{\ee}{\end{eqnarray*}}
\newcommand{\bea}{\begin{eqalph}}
\newcommand{\eea}{\end{eqalph}}
\newcommand{\cA}{{\cal A}}

\newcommand{\cM}{{\cal M}}

\newcommand{\cC}{{\cal C}}

\newcommand{\cH}{{\cal H}}

\newcommand{\al}{\alpha}

\newcommand{\dl}{\delta}
\newcommand{\la}{\lambda}

\newcommand{\f}{\phi}

\newcommand{\Om}{\Omega}
\newcommand{\m}{\mu}

\newcommand{\g}{\gamma}

\newcommand{\vt}{\vartheta}

\newcommand{\up}{\upsilon}
\newcommand{\lng}{\langle}
\newcommand{\rng}{\rangle}

\newcommand{\w}{\wedge}

\newcommand{\wh}{\widehat}
\newcommand{\ol}{\overline}
\newcommand{\dr}{\partial}

\newcommand{\ot}{\otimes}

\newcommand{\ve}{\varepsilon}

\let\ssection=\section
\renewcommand{\section}{\setcounter{equation}{0}\ssection}

\newcounter{eqalph}
\newcounter{equationa}
\newcounter{remark}
\newcounter{example}
\newcounter{theorem}
\newcounter{proposition}
\newcounter{lemma}
\newcounter{corollary}
\newcounter{definition}
\newenvironment{eqalph}{\stepcounter{equation}
\setcounter{equationa}{\value{equation}}
\setcounter{equation}{0}

\begin{eqnarray}}{\end{eqnarray}\setcounter{equation}{\value{equationa}}}

\def\theremark{\arabic{remark}}
\def\therexample{\arabic{remark}}

\def\thetheorem{\arabic{theorem}}

\newcommand{\mar}[1]{}

\begin{document}
\hbox{}

{\parindent=0pt

{\Large \bf Quantization of noncommutative completely integrable
Hamiltonian systems}

\bigskip

{\sc G.Giachetta}$^{\rm a}$, {\sc L.Mangiarotti}$^{\rm a}$, {\sc G.
Sardanashvily}$^{\rm b}$

{\small

\medskip

$^{\rm a}$ {\it Department of Mathematics and Informatics, University of
Camerino, 62032 Camerino (MC), Italy}

$^{\rm b}$ {\it Department of Theoretical Physics,
Moscow State University, 117234 Moscow, Russia}

}

}

\bigskip
\bigskip

{\small

\noindent {\bf Abstract.} Integrals of motion of a Hamiltonian
system need not commute. The classical Mishchenko--Fomenko theorem
enables one to quantize a noncommutative completely integrable
Hamiltonian system around its invariant submanifold as the abelian
one.

\bigskip


}

\bigskip
\bigskip

Recall that an autonomous Hamiltonian system on a $2n$-dimensional
symplectic manifold $(Z,\Om)$ is called completely integrable
(henceforth CIS) if it admits $n$ independent integrals of motion
$\{H_1,\ldots, H_n\}$ in involution. Let $M$ be its regular
connected invariant submanifold. The classical Liouville--Arnold
theorem [1-3] and its generalization [4,5] for noncompact
invariant submanifolds state that an open neighbourhood $U_M$ of
$M$ can be provided with the action-angle coordinates $(J_a,y^a)$
such that a symplectic form on $U_M$ reads $\Om= dJ_a\w dy^a$, and
the integrals of motion $H_a$ together with a Hamiltonian $\cH$
are expressed only in the action coordinates $(J_a)$.

However, integrals of motion of a Hamiltonian system need not
commute. A Hamiltonian system on a symplectic manifold $(Z,\Om)$
is called a noncommutative CIS if it admits $n\leq k<2n$ integrals
of motion $\{H_1,\ldots,H_k\}$ which obey the following
conditions.

(i) The smooth real functions $H_i$ are independent on $Z$, i.e.,
the $k$-form $\op\w^k dH_i$ nowhere vanishes. Their common level
surfaces are regular invariant submanifolds which make $Z$ into a
fibered manifold
\mar{nc4}\beq
H:Z\to N\subset \Bbb R^k. \label{nc4}
\eeq

(ii)  There exist smooth real functions $s_{ij}: N\to \Bbb R$ such that
the Poisson bracket of integrals of motion reads
\mar{nc1}\beq
\{H_i,H_j\}= s_{ij}\circ H, \qquad i,j=1,\ldots, k, \label{nc1}
\eeq
where the matrix  function $(s_{ij})$ is of constant corank
$m=2n-k$ at all points of $N$.

If $k=n$, we are in the case of an abelian CIS. A noncommutative
CIS is exemplified by a spherical top possessing the Lie algebra
$so(3)$ of three independent integrals of motion on a certain
four-dimensional reduced subspace of the momentum phase space.

Let us additionally assume that the Hamiltonian vector fields
$\vt_i$ of integrals of motion $H_i$ are complete and their
invariant manifolds are connected and mutually diffeomorphic. Then
the classical Mishchenko--Fomenko theorem [6-8]  and its
generalization [9] for noncompact invariant submanifolds state
that every invariant submanifold $M$ is diffeomorphic to a
toroidal cylinder $\Bbb R^{m-r}\times T^r$, $m=2n-k$, coordinated
by $(y^a)$, and it admits an open fibered neighbourhood $H:U_M\to
N_M$ endowed with action-angle coordinates $(J_a,p_A,q^A,y^a)$
such that a symplectic form on $U_M$ reads
\mar{nc3}\beq
\Om= dJ_a\w dy^a + dp_A\w dq^A, \label{nc3}
\eeq
and a Hamiltonian $\cH$ depends only on the action coordinates
$J_a$.

One can say something more. The base $N$ (\ref{nc4}) is provided
with a unique coinduced Poisson structure $\{,\}_N$ of rank $2k-n$
such that $H$ is a Poisson morphism. Furthermore, every invariant
submanifold $M$ is a maximal integral manifold of the involutive
distribution spanned by the Hamiltonian vector fields $\up_a$ of
the pull-back $H^*C_a(z)= C_a(H_i(z))$ onto $U_M$ of $m$
independent Casimir functions $\{C_1,\ldots, C_m\}$ on an open
neighbourhood $N_M$ of the point $H(M)\subset N$.  The original
integrals of motion are smooth functions of coordinates
$(J_a,q^A,p_A)$, but the Casimir functions
\mar{j100}\beq
C_a(H_i(J_b,q^A,p_A))= C_a(J_b) \label{j100}
\eeq
depend only on the  action coordinates $J_a$. Moreover, a
Hamiltonian $\cH(J_b)=\cH(C_a(J_b))$ is expressed in action
variables $J_a$ through the Casimir functions (\ref{j100}).

We aim to quantize a noncommutative CIS written in the
action-angle variables around its invariant submanifold. Since
$(J_a,p_A,q^A)$ are coordinates on $N_M$, they are integrals of
motion which constitute a noncommutative CIS
\mar{j41}\beq
\{J_a, p_A\}=\{J_a,q^A\}=0, \qquad \{p_A,q^B\}=\dl^B_A,
\label{j41}
\eeq
on $U_M$ equivalent to the original one (\ref{nc1}). Furthermore,
this CIS can be treated as a particular abelian CIS possessing $n$
integrals of motion $\{J_a, p_A\}$ and action-angle coordinates
$(J_a, p_A, q^A, y^a)$ on $U_M$, where  $(q^A, y^a)$ are angle
coordinates on its invariant submanifold
\mar{j10}\beq
\cM=V_M\times \Bbb R^{m-r}\times T^r\subset \Bbb R^{n-r}\times
T^r, \label{j10}
\eeq
where $V_M$ is a base of the fibration $U_M\ni (J_a,p_A,q^A)\to
(q^A)\in V_M$. Therefore, the noncommutative CIS (\ref{j41}) can
be quantized as the abelian one. Strictly speaking, this
quantization fails to be a quantization of the original CIS
(\ref{nc1}) because $H_i(J_a,q^A,p_A)$ are not linear functions
and, consequently, the algebras (\ref{nc1}) and (\ref{j41}) are
not isomorphic in general. As a result, one however can obtain the
Hamilton operator $\wh\cH$ and the Casimir operators $\wh C_a$ of
an original CIS and their spectra.

There are different approaches to quantization of abelian CISs
[10-14]. It should be emphasized that action-angle coordinates
need not be globally defined on the momentum phase space of a CIS,
but form an algebra of Poisson canonical commutation relations on
an open neigbourhood $U_M$ of an invariant submanifold $M$.
Therefore, quantization of a CIS with respect to the action-angle
variables is a quantization of the Poisson algebra $C^\infty(U_M)$
of real smooth functions on $U_M$. A key point is that, since
$U_M$ is not a contractible manifold, the geometric quantization
technique should be called into play in order to quantize a CIS
around its invariant submanifold. Geometric quantization of
abelian CISs has been studied at first with respect to the
polarization spanned by Hamiltonian vector fields of integrals of
motion [11,15]. For example, the well-known  Simms quantization of
a harmonic oscillator is of this type. However, one meets a
problem that the associated quantum algebra contains affine
functions of angle coordinates on a torus which are ill defined.
As a consequence, elements of the carrier space of this
quantization fail to be smooth, but are tempered distributions. We
have developed a different variant of geometric quantization of
abelian CISs [14,16-17]. Since a Hamiltonian of a CIS depends only
on action variables, it seems natural to provide the Schr\"odinger
representation of action variables by first order differential
operators on functions of angle coordinates. For this purpose, one
should choose the angle polarization of a symplectic manifold
spanned by almost-Hamiltonian vector fields of angle variables.
This quantization scheme is straightforwardly extended to the case
of a noncompact invariant submanifold (\ref{j10}). Since the
action-angle coordinates $(J_a, p_A, q^A,  y^a)$ are canonical for
the symplectic form $\Om$ (\ref{nc3}), geometric quantization of
the symplectic annulus $(U_M,\Om)$ in fact is equivalent to
geometric quantization of the cotangent  bundle $T^*\cM$ of the
toroidal cylinder $\cM$ (\ref{j10}) endowed with the canonical
symplectic form $\Om$ (\ref{nc3}). In this case, the  above
mentioned angle polarization coincides with the vertical tangent
bundle $VT^*\cM$ of $T^*\cM\to \cM$.

Let $(q^A,y^i,\al^\m))$ be coordinates on the toroidal cylinder
(\ref{j10}), where $(\al^1,\ldots, \al^r)$ are angle coordinates
on a torus $T^r$, and let $(p_A, J_i, J_\m)$ be the corresponding
action coordinates (i.e., the induced fibered coordinates on
$T^*\cM$).  Since the symplectic form $\Om$ (\ref{nc3}) is exact,
the quantum bundle is defined as a trivial complex line bundle
$\cC$ over $T^*\cM$. Let its trivialization hold fixed. Any other
trivialization leads to an equivalent quantization of $T^*\cM$.
Given the associated fiber coordinate $c\in\Bbb C$ on $\cC\to
T^*\cM$, one can treat its sections as smooth complex functions on
$T^*\cM$.

The Konstant--Souriau prequantization formula associates to
every smooth real function $f$ on
$T^*\cM$ the first order differential operator
\mar{lqq46}\beq
\wh f=-i\nabla_{\vt_f} + f \label{lqq46}
\eeq
on sections of $\cC\to T^*\cM$, where $\vt_f$ is the Hamiltonian
vector field of $f$ and $\nabla$ is the covariant differential
with respect to a suitable $U(1)$-principal connection $A$ on
$\cC$. This connection preserves the Hermitian metric
$g(c,c')=c\ol c'$ on $\cC$, and its curvature obeys the
prequantization condition $R=i\Om$. It reads
\mar{ci20}\beq
A=A_0 +ic(p_Adq^A + J_jdy^j + J_\m d\al^\m)\ot\dr_c, \label{ci20}
\eeq
where $A_0$ is a flat $U(1)$-principal connection on $\cC\to
T^*\cM$. The classes of gauge nonconjugated flat
principal connections on $\cC$ are indexed
by the set $\Bbb R^r/\Bbb Z^r$ of homomorphisms of the de Rham cohomology
group
\be
H^1(T^*\cM)=H^1(\cM)=H^1(T^r)=\Bbb R^r
\ee
of $T^*\cM$ to $U(1)$.
We choose their representatives of the form
\be
&& A_0[(\la_\m)]=dp_A\ot\dr^A + dJ_j\ot\dr^j + dJ_\m\ot\dr^\m+
dq^A\ot\dr_A +dy^j\ot\dr_j + \\
&&\qquad d\al^\m\ot(\dr_\m +i\la_\m c\dr_c),
\qquad \la_\m\in [0,1).
\ee
Accordingly, the relevant connection (\ref{ci20}) on $\cC$ reads
\mar{ci14}\ben
&& A[(\la_\m)]= dp_A\ot\dr^A + dJ_j\ot\dr^j + dJ_\m\ot\dr^\m
+\label{ci14}\\
&& \qquad dq^A\ot(\dr_A + ip_Ac\dr_c) + dy^j\ot(\dr_j + iJ_jc\dr_c)
+ d\al^\m\ot(\dr_\m + i(J_\m +\la_\m)c\dr_c). \nonumber
\een
For the sake of simplicity, we further assume that the numbers
$\la_\m$ in the expression (\ref{ci14}) belong to $\Bbb R$, but
bear in mind that connections $A[(\la_\m)]$ and $A[(\la'_\m)]$
with $\la_\m-\la'_\m\in\Bbb Z$ are gauge conjugated.

Let us choose the above mentioned angle polarization $VT^*\cM$. Then
the corresponding quantum algebra
$\cA$ of $T^*\cM$ consists of affine functions
\be
f=a^A(q^B,y^j,\al^\nu)p_A + a^i(q^B,y^j,\al^\nu)J_i +
a^\m(q^B,y^j,\al^\nu)J_\m + b(q^B,y^j,\al^\nu)
\ee
in action coordinates $(p_A, J_i, J_\m)$. Given a connection
(\ref{ci14}), the corresponding operators (\ref{lqq46}) read
\mar{lmp135}\beq
\wh f=(-ia^A\dr_A-\frac{i}{2}\dr_Aa^A)
+(-ia^i\dr_i-\frac{i}{2}\dr_ia^i) +
(-ia^\m\dr_\m-\frac{i}{2}\dr_\m a^\m-a^\m\la_\m)+
b. \label{lmp135}
\eeq
They are
self-adjoint operators in the pre-Hilbert
space $\Bbb C^\infty_c(\cM)$
of smooth complex functions of compact support on $\cM$ endowed
with the Hermitian form
\be
\lng \psi|\psi'\rng=\left(\frac1{2\pi}\right)^r\op\int_{\cM} \psi \ol
\psi' d^{n-m} q d^{m-r} y d^r\al,
\qquad \psi,\psi'\in \Bbb C^\infty_c(\cM).
\ee
Note that any function $\psi\in \Bbb C^\infty_c(\cM)$ is expanded into
the series
\mar{j25}\beq
\psi= \op\sum_{(n_\m)} \f(q^B, y^j)_{(n_\m)}\exp[in_\m\al^\m], \qquad
(n_\m)=(n_1,\ldots,n_r)\in\Bbb Z^r, \label{j25}
\eeq
where $\f(q^B, y^j)_{(n_\m)}$ are functions of compact support on $\Bbb
R^{n-r}$.
In particular, the action operators (\ref{lmp135}) read
\mar{j11}\beq
\wh p_A=-i\dr_A, \qquad \wh J_j=-i\dr_j, \qquad \wh J_\m=-i\dr_\m -\la_\m.
\label{j11}
\eeq
It should be emphasized that
\mar{j31}\beq
\wh a \wh p_A\neq \wh{ap_A}, \qquad \wh a \wh J_j\neq \wh{a J_j}, \qquad
\wh a \wh J_\m\neq \wh{aJ_\m}, \qquad a\in C^\infty(\cM). \label{j31}
\eeq

The operators (\ref{lmp135}) provide the desired quantization of a
noncommutative CIS written with respect to the action-angle
coordinates. They satisfy the Dirac condition
\mar{j45}\beq
[\wh f,\wh f']=-i\wh{\{f,f'\}}, \qquad f,f'\in\cA. \label{j45}
\eeq
However, both a Hamiltonian $\cH$ and original integrals of motion
$H_i$ do not belong to the quantum algebra $\cA$, unless they are
affine functions in the action coordinates $(p_A, J_i, J_\m)$. It
is a well-known problem of the Schr\"odinger representation. In
some particular cases, integrals of motion $H_i$ can be
represented by differential operators, but this representation
fails to be unique because of inequalities (\ref{j31}), and the
Dirac condition (\ref{j45}) need not be satisfied. At the same
time, both a Hamiltonian $\cH$ and the Casimir functions $C_\la$
depend only on action variables $J_i,J_\m$. If they are polynomial
in $J_i$, one can associate to them the operators $\wh\cH=\cH(\wh
J_i,\wh J_\m)$, $\wh C_\la=C_\la(\wh J_i,\wh J_\m)$ acting in the
space $\Bbb C^\infty_c(\cM)$ by the law
\be
&& \wh \cH\psi= \op\sum_{(n_\m)} \cH(\wh
J_i,n_\m-\la_\m)\f(q^A,y^j)_{(n_\m)}
\exp[in_\m\al^\m], \\
&& \wh C_\la\psi= \op\sum_{(n_\m)} C_\la(\wh
J_i,n_\m-\la_\m)\f(q^A,y^j)_{(n_\m)} \exp[in_\m\al^\m].
\ee

Let us mention a particular class of CISs
whose integrals of motion $\{H_1,\ldots,H_k\}$ form a $k$-dimensional real
Lie algebra $\cG$ of rank $m$ with the commutation relations
\be
\{H_i,H_j\}= c_{ij}^h H_h, \qquad c_{ij}^h={\rm const.}
\ee
In this case, nonvanishing complete Hamiltonian vector fields
$\vt_i$ of $H_i$ define a free Hamiltonian action on $Z$ of some
connected Lie group $G$ whose Lie algebra is isomorphic to $\cG$.
Orbits of $G$ coincide with $k$-dimensional maximal integral
manifolds of the regular distribution on $Z$ spanned by
Hamiltonian vector fields $\vt_i$ [19]. Furthermore, one can treat
$H$ (\ref{nc4}) as an equivariant momentum mapping of $Z$ to the
Lie coalgebra $\cG^*$, provided with the coordinates
$x_i(H(z))=H_i(z)$, $z\in Z$ [18,20]. In this case, the coinduced
Poisson structure $\{,\}_N$ on the base $N$ coincides with the
canonical Lie--Poisson structure on $\cG^*$ given by the Poisson
bivector field
\be
w=\frac12 c_{ij}^h x_h\dr^i\w\dr^j.
\ee
Recall that the coadjoint action of $\cG$ on $\cG^*$ reads
\mar{j102}\beq
\ve_i(x_j)=c_{ij}^hx_h. \label{j102}
\eeq
Casimir functions of the Lie--Poisson structure are exactly the
coadjoint invariant functions on $\cG^*$. They are constant on
orbits of the coadjoint action of $G$ on $\cG^*$. Given a point
$z\in Z$ and the orbit $G_z$ of $G$ in $Z$ through $z$, the
fibration $H$ (\ref{nc4}) projects this orbit onto the orbit
$G_{H(z)}$ of the coadjoint action of $G$ in $\cG^*$ through
$H(z)$. Moreover, the inverse image $H^{-1}(G_{H(z)})$ of
$G_{H(z)}$ coincides with the orbit $G_z$. It follows that any
orbit of $G$ in $Z$ is fibered in invariant submanifolds.

The Mishchenko--Fomenko
theorem has been mainly applied to CISs whose integrals of motion form a
compact Lie algebra. The group $G$ generated by flows of
their Hamiltonian vector fields is compact, and every orbit of $G$ in $Z$
is compact. Since a fibration of a compact manifold possesses compact
fibers, any invariant submanifold of such a noncommutative CIS is
compact.

For instance, let us consider the above mentioned noncommutative
CIS with the Lie algebra $\cG=so(3)$ of integrals of motion
$\{H_1,H_2,H_3\}$ on a four-dimensional symplectic manifold
$(Z,\Om)$, namely,
\mar{j50}\beq
\{H_1,H_2\}=H_3, \qquad \{H_2,H_3\}=H_1, \qquad \{H_3,H_1\}=H_2.
\label{j50}
\eeq
The rank of this Lie algebra equals one. Since it is compact, an
invariant submanifold of a CIS in question is a circle $M=S^1$. We
have a fibered manifold $H:Z\to N$ onto an open subset $N\subset
\cG^*$ of the Lie coalgebra $\cG^*$. This fibered manifold is a
fiber bundle since its fibers are compact [21]. The base $N$ is
endowed with the coordinates $(x_1,x_2,x_3)$ such that which
integrals of motion $\{H_1,H_2,H_3\}$ on $Z$ read
\be
H_1=x_1, \qquad H_2=x_2, \qquad H_3=x_3.
\ee
As was mentioned above, the coinduced Poisson structure on $N$
is the Lie--Poisson structure
\mar{j51}\beq
w= x_2\dr^3\w\dr^1 + x_3\dr^1\w\dr^2 + x_1\dr^2\w\dr^3. \label{j51}
\eeq
The coadjoint action (\ref{j102}) of $so(3)$ reads
\be
\ve_1=x_3\dr^2-x_2\dr^3, \qquad \ve_2=x_1\dr^3-x_3\dr^1, \qquad
\ve_3=x_2\dr^1-x_1\dr^2.
\ee
An orbit of the coadjoint action of dimension 2 is given by the
equation
\be
(x_1^2 + x_2^2 + x_3^2)={\rm const}.
\ee

Let $M$ be an invariant submanifold such that the point $H(M)\in
\cG^*$ belongs to an orbit of the coadjoint action of maximal
dimension 2. Let us consider an open fibered neighbourhood
$U_M=N_M\times S^1$ of $M$ which is a trivial bundle over an open
contractible neighbourhood $N_M$ of $H(M)$ endowed with the
coordinates $(r,x_1,\g)$ defined by the equalities
\mar{j52}\beq
r=(x_1^2 + x_2^2 + x_3^2)^{1/2}, \quad
x_2=(r^2-x_1^2)^{1/2}\sin\g, \quad x_3=(r^2-x_1^2)^{1/2}\cos\g.
\label{j52}
\eeq
Here, $r$ is a Casimir function on $\cG^*$. It is readily observed
that the coordinates (\ref{j52}) are the Darboux coordinates of
the Lie--Poisson structure (\ref{j51}) on $N_M$, namely,
\mar{j53}\beq
w=\frac{\dr}{\dr x_1}\w \frac{\dr}{\dr \g}. \label{j53}
\eeq
Let $\vt_r$ be the Hamiltonian vector field of the Casimir
function $r$ (\ref{j52}). It is a combination
\be
\vt_r=\frac1{r}(x_1\vt_1 + x_2\vt_2 + x_3\vt_3)
\ee
of the Hamiltonian vector fields $\vt_i$ of integrals of motion
$H_i$. Its flows are invariant submanifolds. Let $\al$ be a
parameter along the flows of this vector field, i.e.,
\be
\vt_r= \frac{\dr}{\dr \al}.
\ee
Then $U_M$ is provided with the action-angle coordinates
$(r,x_1,\g,\al)$ such that the Poisson bivector associated to the
symplectic form $\Om$ on $U_M$ reads
\mar{j54}\beq
W= \frac{\dr}{\dr r}\w \frac{\dr}{\dr \al} + \frac{\dr}{\dr x_1}\w
\frac{\dr}{\dr \g}. \label{j54}
\eeq
Accordingly, Hamiltonian vector fields of integrals of motion take
the form
\be
&& \vt_1= \frac{\dr}{\dr \g}, \\
&& \vt_2= r(r^2-x_1^2)^{-1/2}\sin\g
\frac{\dr}{\dr \al} - x_1 (r^2-x_1^2)^{-1/2}\sin\g
\frac{\dr}{\dr \g} - (r^2-x_1^2)^{1/2}\cos\g
\frac{\dr}{\dr x_1}, \\
&& \vt_3= r(r^2-x_1^2)^{-1/2}\cos\g
\frac{\dr}{\dr \al} - x_1 (r^2-x_1^2)^{-1/2}\cos\g
\frac{\dr}{\dr \g} + (r^2-x_1^2)^{1/2}\sin\g
\frac{\dr}{\dr x_1}, \\
&& \vt_1\w\vt_2\w\vt_3= r \frac{\dr}{\dr \al}\w\frac{\dr}{\dr \g} \w
\frac{\dr}{\dr x_1}\neq 0.
\ee

The action-angle variables $\{r, H_1=x_1, \g\}$ constitute a
noncommutative CIS
\mar{j55}\beq
\{r, H_1\}=0, \qquad \{r, \g\}=0, \qquad \{H_1,\g\}=1, \label{j55}
\eeq
on $U_M$. This noncommutative CIS is related to the original one
by the transformations
\be
r=(H_1^2 + H_2^2 + H_3^2)^{1/2}, \qquad  H_2=(r^2-H_1^2)^{1/2}\sin\g,
\qquad H_3=(r^2-H_1^2)^{1/2}\cos\g.
\ee
Its Hamiltonian is expressed only in the action
variable $r$.

Let us quantize the noncommutative CIS (\ref{j55}). We obtain the
algebra of operators
\be
\wh f= a(-i\frac{\dr}{\dr \al} -\la) -ib\frac{\dr}{\dr \g}
-\frac{i}2(\frac{\dr a}{\dr \al} + \frac{\dr b}{\dr \g}) + c,
\ee
where $a$, $b$, $c$ are smooth functions of angle coordinates $(\g,\al)$
on the cylinder $\Bbb R\times S^1$. In particular, the action operators
read
\mar{j56}\beq
\wh r= -i\frac{\dr}{\dr \al} -\la, \qquad \wh H_1=-i\frac{\dr}{\dr \g}.
\label{j56}
\eeq
These operators act in the space of smooth complex functions
\be
\psi(\g,\al)= \op\sum_k \f(\g)_k\exp[ik\al]
\ee
of compact support on $\Bbb R\times S^1$. A Hamiltonian $\cH(r)$ of a
classical CIS can also be represented by the operator
\be
\wh \cH(r)\psi=\op\sum_k \cH(k-\la)\f(\g)_k\exp[ik\al]
\ee
on this space.

For instance, let us consider a spherical top whose integrals of motion
$\{H_1, H_2, H_3\}$ are angular momenta, and a Hamiltonian reads
\be
\cH=\frac12 I(H_1^2 + H_2^2 + H_3^2)= \frac12 I r^2,
\ee
where $I$ is a rotational constant. The momentum phase space of a
spherical top is the cotangent bundle $Z'=T^*RP^3$ of the group
space $RP^3$ of $SO(3)$. It is a trivial bundle $Z'=RP^3\times
\cG^*$ provided with the symplectic structure given by the
non-degenerate Poisson bracket
\be
\{x_i,x_j\}= c_{ij}^hx_h, \qquad \{\al^i,\al^j\}=0, \qquad
\{x_j,\al^i\}=\dl^i_j,
\ee
where $\al^i$ are group parameters. Note that it is not the
canonical symplectic structure on the cotangent bundle. Let us
consider a four-dimensional submanifold $Z\subset Z'$ of points
which belong to the one-dimensional trajectories of a spherical
top passing through the unit of $SO(3)$. These trajectories are
exactly the invariant submanifolds of the noncommutative CIS
(\ref{j50}), and $Z$ is the corresponding fibered manifold $H:Z\to
N=\cG^*\setminus \{0\}$. This fibered manifold is not trivial. In
particular, the restriction of $Z$ to a coadjoint orbit $r=$const.
of $N$ is a nontrivial fiber bundle $SO(3)=RP^3\to
SO(3)/SO(2)=S^2$. Its restriction to a cycle $S^1$, $r=$const.,
$x_1=$const., is isomorphic to the trivial bundle $T^2\to S^1$.
However, the parameter $\al$ along the flows of the Hamiltonian
vector field $\vt_r$ need not perform such a trivialization.
Therefore, the action-angle coordinate chart $(r,x_1,\g,\al)$ is
defined on an open neighbourhood $U_M=N_M\times S^1$ of an
invariant submanifold $M$ where $N_M$ is an open contractible
neighbourhood of $H(M)$ diffeomorphic to $\Bbb R^3$.

A familiar quantization of a spherical top in fact reduces to a
linear representation of the Lie algebra $so(3)$ by differential
operators $\{\wh H_1, \wh H_2, \wh H_3\}$ in the space of smooth
complex functions on a sphere $S^2$. In comparison with this
quantization, the operators (\ref{j56}) provide a representation
of the algebra  of canonical commutation relations (\ref{j55})
(but not the Lie algebra $so(3)$) in the space of smooth complex
functions of compact support on $\Bbb R\times S^1$.

\end{document}